\documentclass[cameraready]{Interspeech}
\title{Tracking the Trend in How Speech Synthesizers Deceive People}

\author[orcid=0009-0004-9604-168X, equalcontribution]{Milan}{Šalko}
\author[orcid=0000-0002-4717-1910, equalcontribution]{Anton}{Firc}
\author[orcid=0000-0002-9009-2193, equalcontribution, correspondingauthor]{Kamil}{Malinka}
\author[orcid=0009-0000-5722-0571]{\\Vojtěch}{Staněk}
\author[orcid=0000-0002-2875-9567]{Martin}{Perešini}
\author[orcid=0009-0000-1238-4062]{Filip}{Pleško}
\author[orcid=0009-0004-6055-5136]{Jakub}{Reš}
\address{
    Security@FIT, Brno University of Technology, Czech Republic
}

\email{\{isalko, ifirc, malinka, istanek, iperesini, iplesko, iresj\}@fit.vut.cz}


\usepackage{comment}
\usepackage{graphicx}
\usepackage{subcaption}
\usepackage{adjustbox}
\usepackage{multirow}

\usepackage{booktabs} 
\usepackage[table]{xcolor} 
\usepackage{pgf}

\usepackage{booktabs}
\usepackage[table]{xcolor}

\newcommand{\score}[1]{%
  \ifdim #1pt<0.20pt
    \cellcolor{red!80}{#1}%
  \else\ifdim #1pt<0.40pt
    \cellcolor{red!45}{#1}%
  \else\ifdim #1pt<0.60pt
    \cellcolor{orange!35}{#1}%
  \else\ifdim #1pt<0.80pt
    \cellcolor{yellow!45}{#1}%
  \else
    \cellcolor{green!45}{#1}%
  \fi\fi\fi\fi
}

\ifcameraready
  \newcommand{\event}{DevConf\xspace}
\else
  \newcommand{\event}{[Event]\xspace}
\fi

\begin{document}

\maketitle

\begin{abstract}
   Advances in speech synthesis have made deepfake audio highly realistic. Earlier studies reported 70--80\% human detection accuracy, but relied primarily on older synthesizers. We compare human detection for three selected voice synthesis tools released in 2019, 2022, and 2024 with 82 IT professionals, and benchmark humans against six pretrained detectors on the same material. For fully synthetic speech (full spoofs), the F1 score drops from about 90\% for RTVC and YourTTS to 48\% for ElevenLabs, although listeners were explicitly warned that deepfakes were present. For partial spoofing, where only one sentence of an utterance is altered, strict accuracy falls to 9\%, and listeners classify the synthetic sentence as bona fide 77\% of the time. Humans and detectors fail in complementary ways, and neither reliably localizes short manipulations. Additionally, listeners increasingly mislabel bona fide speech as fake, eroding trust in unmanipulated audio. These findings show that human perception alone is unreliable for the selected modern and partial-spoof conditions and motivate procedural verification, provenance, watermarking, and segment-level detection.
\end{abstract}

\section{Introduction}

In recent years, voice deepfake technology has advanced significantly, increasing the risk of misuse in disinformation, fraud, and blackmail~\cite{The_Guardian_2024}. The human ability to distinguish synthetic voices from bona fide speech is crucial, as it affects not only individuals who may fall victim to fraud schemes but also society at large~\cite{prudky2023recognize, deepfakeSurvey}.

The rapid pace of development in synthetic voice technology makes this threat particularly critical, as human vulnerability may change quickly with each new generation of synthesis tools. Current research often lags behind these advances and fails to address emerging threat models, such as \emph{partial spoofing}~\cite{zhang2023partialspoof}, where only short segments of recordings are manipulated~\cite{8999436}.  This raises the need to assess human vulnerability as it evolves with technology, not only to understand current risk levels, but also to inform appropriate security measures. Establishing such a baseline is crucial: without it, organizations cannot perform accurate threat modeling, risk analysis, or select effective mitigation strategies~\cite{Torten2018}.

While the speech-security community has invested heavily in automated countermeasures against synthetic speech (e.g., the ASVspoof and PartialSpoof initiatives~\cite{zhang2023partialspoof,WANG2020101114}), the human listener remains the last line of defense whenever such systems are absent, bypassed, or not yet deployed, for instance, during everyday phone calls, voice messages, or media consumption. Effective security, therefore, requires continuous monitoring of how attacks evolve~\cite{NISTCSF2024}, including how advances in speech synthesis erode the reliability of human judgment~\cite{Malinka2024Comprehensive}.

This study is a questionnaire-based survey in which participants judge the authenticity of utterances. Participants are IT professionals, a hard-to-reach group with above-average security awareness and technical literacy~\cite{Torten2018, Mersinas2015}. The study therefore characterizes performance in a technically literate population, but does not establish an upper bound for the general public.

That humans are unreliable at detecting synthetic speech is, by now, well established: a recent meta-analysis of 56 studies finds that human accuracy is only moderately above chance and depends heavily on the particular audio samples tested~\cite{diel2024human}. We therefore take this baseline as established and instead ask what changes at two points that prior work has \emph{not} examined on the human side: the arrival of modern commercial synthesis and partial spoofing. Earlier work reported that warning listeners that deepfakes may be present improves their accuracy~\cite{Malinka2024Comprehensive}. In our explicitly warned condition, detection of the selected ElevenLabs samples nevertheless remained low. Because we did not include an unwarned control group, we cannot estimate the effect of the warning itself.

In this paper, we present a controlled empirical comparison of human susceptibility to voice deepfakes for three selected synthesis systems released between 2019 and 2024, evaluating both fully synthetic and partially synthetic speech and benchmarking humans against automated detectors on the same material. Our study updates and extends prior work~\cite{Malinka2024Comprehensive, watson2021audio, muller_human_2022, maiPaper, barrington2025people} for a more modern threat landscape. The findings are concerning: human perception alone is becoming unreliable, as modern synthesizers both escape detection and undermine trust in bona fide audio.

\noindent
\textbf{The main contributions} of this paper are:

\begin{itemize}[topsep=0pt, itemsep=0pt]
    \item A systematic, controlled comparison of \emph{human} voice-deepfake detection for three selected synthesizers released between 2019 and 2024, indicating that recognition of fully synthetic speech is substantially lower for the selected commercial tool, ElevenLabs (F1 score 0.48), even when listeners are explicitly warned that deepfakes are present.

    \item One of the first studies of \emph{human} perception of \emph{partial} spoofing, in which a single sentence is replaced within an otherwise bona fide utterance. We show that it is more deceptive than full spoofing (with strict accuracy below 10\%) and provide an analysis and explanation of this phenomenon.

    \item A head-to-head benchmark of humans against six pretrained automated detectors on the same material, revealing that the selected detectors also break down on modern and partially spoofed speech, signalling an emerging erosion of trust in bona fide audio.
\end{itemize}

\subsection{Related work}

Research on human recognition of deepfakes shows inconsistent results. Mai et al.~\cite{maiPaper} measured a 73\% detection success rate with no difference between English and Mandarin. Wang et al.~\cite{WANG2020101114} confirmed users' ability to identify deepfakes in a banking scenario, although they did not report the exact accuracy. Müller et al.~\cite{muller_human_2022} reported an 80\% human detection success rate, particularly for TTS models with native speakers. Watson et al.~\cite{watson2021audio} report accuracy ranging from 42\% to 90\% depending on sentence length and complexity.
Malinka et al.~\cite{Malinka2024Comprehensive} simulated a more realistic attack scenario in which participants were not informed that they might encounter deepfakes, resulting in a dramatic drop in detection accuracy to 3.2\% and highlighting the critical role of prior expectations in human recognition performance. At scale, Warren et al.~\cite{warren2024humans} evaluated humans as audio-deepfake detectors and likewise report degraded performance against modern clones. Most recently, Barrington et al.~\cite{barrington2025people} found that listeners are near chance against modern AI voice clones. Two regularities emerge: reported accuracy is high for older synthesizers (around 73 to 90\%) but drops sharply for the most recent ones, and, crucially, all of these studies evaluate only \emph{fully} synthetic speech, with most relying on listeners being primed that deepfakes are present.

While previous studies assessed human ability to detect deepfake voices, they typically focused on individual synthesizers or fully synthetic audio. Our study addresses two critical gaps in the literature. First, we systematically compare human detection performance for three selected speech synthesis tools released between 2019 and 2024, revealing substantial system-specific differences in recognition accuracy within the same user group. Second, we introduce and evaluate partial spoofing, an underexplored manipulation technique in which only a short segment of a bona fide recording is synthetically altered. This hybrid form of deepfake audio accurately reflects realistic scenarios of misinformation and deception, where attackers seek to manipulate public perception or trust by altering parts of the bona fide content. The \emph{automated} detection and localization of such manipulations has begun to be addressed by the PartialSpoof database and countermeasures~\cite{zhang2023partialspoof}, by neural speech-editing detection~\cite{zhang2025partialedit}, by spoof diarization~\cite{zhang2024spoofdiar}, and by temporal forgery localization~\cite{wu2024coarse}; large-scale studies have also measured human detection of \emph{full} spoofs~\cite{warren2024humans}. To our knowledge, however, whether \emph{human listeners} can detect or localize a single synthetic sentence embedded in bona fide speech has not been studied; this human counterpart to partial spoofing is the gap we fill. Our study shows that partial spoofing significantly impairs human detection and increases false positives, highlighting an urgent need for new detection methods and resilience strategies.

Unlike previous studies, which primarily target the general public, we evaluate IT professionals, a population with above-average technological literacy and security awareness~\cite{Torten2018,Mersinas2015}. This deliberate choice allows us to examine a technically literate population, but IT expertise does not necessarily imply expertise in speech forensics, and the results may not generalize to lay listeners.

\section{Experiment design}

\iffalse
The proposed experiment aims to analyze trends in people's ability to recognize voice deepfakes generated by different generations of speech synthesizers. It also investigates how detection accuracy changes when a deepfake is embedded within bona fide speech. The experiment is structured as a questionnaire-based survey conducted among a selected sample of IT professionals. It consists of two parts: the first part focuses on fully synthetic (\textit{full spoof}) recordings, where speech synthesis models generate entire recordings (comprising four sentences). This section seeks to address the following research question:
\fi

The experiment is a two-part questionnaire-based survey. The first uses fully synthetic (\textit{full spoof}) recordings, in which all four sentences are generated, and addresses the research question:

\noindent \textbf{RQ1:} \textit{How does human voice-deepfake detection performance differ among the selected speech synthesizers?}

The synthetic recordings are created with three widely used deepfake tools released roughly two years apart, allowing us to compare human recognition across the selected systems.

\iffalse
This part analyzes how people's ability to recognize deepfake voices changes over time in the context of advancements in deepfake synthesizers. The synthetic recordings are created using different generations of deepfake tools, released approximately two years apart, each of which was one of the most widely used methods for generating deepfakes at the time. This approach allows us to track the impact of improving synthetic voice quality on people's ability to recognize deepfakes.
\fi

The second part works only with a partially altered (\textit{partial spoof}) recording where only one sentence out of four is replaced in the recording. This part aims to answer the following research question: 

\noindent \textbf{RQ2:} \textit{How does the human ability to recognize voice deepfake change when deepfakes are embedded between bona fide recordings?}

This setting matters for misinformation: an attacker need not fabricate an entire message, since changing a single sentence can alter its meaning, distorting public opinion or damaging reputations. We therefore use partial-spoof recordings, where most of the audio is taken from a bona fide interview and one sentence is replaced by synthetic speech, to measure how well listeners detect such modifications in a realistic context.

Finally, we benchmark human judgments against modern automated deepfake speech detectors evaluated on the same material, addressing:

\noindent \textbf{RQ3:} \textit{How does human voice deepfake detection compare with modern automated detectors on the same recordings?}

This comparison matters because automated countermeasures are often treated as the technical answer to synthetic speech, yet their robustness under modern commercial synthesis and partial-spoof conditions is uncertain. Evaluating humans and detectors on identical recordings reveals whether detectors provide a safety margin beyond human perception, or whether both fail under the same emerging threat models.

\subsection{Used speech synthesis tools}
For the experiment, we used three speech synthesis tools released in different periods: Real-Time Voice Cloning (2019), YourTTS (2022), and ElevenLabs (2024). ElevenLabs is widely available, achieves high speech quality, and is the only commercially available system in our selection. Because each release period is represented by one system and ElevenLabs is also the only commercial system, our results support comparisons among these selected tools but do not establish a general temporal trend in speech synthesis.

\textit{Real-Time Voice Cloning (RTVC)}~\cite{rtvc} from 2019 uses Tacotron2~\cite{tacotron} and WaveRNN~\cite{WAVE} for fast speech synthesis and can clone voice using only very short segments of target speech (10 seconds).

\textit{YourTTS} (2022)~\cite{yourtts} is a multilingual zero-shot model built on the VITS~\cite{kim2021conditionalvariationalautoencoderadversarial} architecture, which enables high-fidelity voice cloning and speech synthesis.

The latest model \textit{ElevenLabs} (2024)~\cite{elevenlabs} uses advanced neural networks to generate speech with natural intonation and emotional variability. We select this model for its wide availability and ability to mimic a variety of voices realistically. It is the only one of the selected models that is commercially available and requires payment.

We used pre-trained weights to generate recordings for all three tools. All the tools were used in a zero-shot manner. Thus, no additional training or fine-tuning of the target speaker was performed. The source speech used for zero-shot synthesis was 3 minutes long.

\subsection{Speaker selection for deepfake creation} 

In selecting speakers for our experiments, we decided to select well-known persons, as their voices better simulate a realistic scenario in which one encounters deepfake misinformation from a public figure. We thus chose celebrities who speak English to create deepfake voices. We focused on well-known figures from the entertainment industry. We deliberately avoided politicians to minimize possible sympathy or antipathy towards the individuals, even though their voices are more frequently targeted by deepfake manipulation.

The audio material for the experiment came from YouTube interviews. This selection from publicly available sources best simulates the attack vector~\cite{FITPUB12595}. A total of nine celebrities (4 men, 5 women) were selected. The selected celebrities are Adele (AD), Miley Cyrus (MC), Liam Neeson (LN), Selena Gomez (SG), Tom Holland (TH), Jennifer Aniston (JA), Jeff Bridges (JB), Cate Blanchett (CB), and Morgan Freeman (MF).

\subsection{Survey structure and evaluation protocol}
\label{survey_structure}
\begin{figure}[t]
    \centering
    \includegraphics[width=0.8\linewidth]{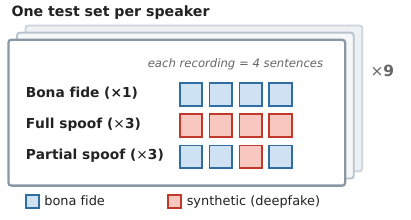}
    \caption{Structure of one of the nine speaker test sets. Each recording comprises four sentence-level segments that participants judged independently as bona fide or synthetic.}
    \label{fig:schematic}
\end{figure}

The survey consisted of nine test sets, each corresponding to one selected speaker. Each set contained seven recordings, with each recording consisting of four parts, each representing one sentence. Specifically, each set included one bona fide recording, three full-spoof recordings in which all four sentences were synthetically generated, and three partial-spoof recordings in which only one sentence was replaced with a deepfake, while the remaining three sentences were bona fide. Figure~\ref{fig:schematic} illustrates the composition of a single test set. Consequently, six of the seven recordings in each speaker set contained some synthetic speech. This deliberately high and fixed prevalence enabled balanced comparisons across spoofing conditions, but it does not reflect real-world prevalence and may have increased participants' tendency to classify speech as synthetic.

In the partial-spoof recordings, one randomly selected sentence was replaced with a deepfake generated using one of the speech synthesis tools described above, while the remaining three sentences were taken from a bona fide recording of the speaker. This setup reflects how even a single manipulated sentence can distort the meaning of an entire message, making the deception more convincing and harder to detect.

The content for both full-spoof and partial-spoof recordings was generated using ChatGPT-4. We prompted the model to create conversational segments that the selected speakers could plausibly say. These generated texts were then converted to audio using the selected speech synthesis tools. For partial-spoof recordings, we provided ChatGPT-4 with the original transcript and instructed it to modify one sentence. The altered sentence was then synthesized using the selected deepfake tools and integrated into the original recording. During post-processing, we normalized volume levels and reduced background noise to ensure that the modified recordings sounded natural and free of distracting artifacts. We used generated replacement text to model a misinformation scenario in which a plausible synthetic sentence changes the meaning of an otherwise bona fide statement. Consequently, partial-spoof performance may reflect both acoustic and linguistic differences between the replaced sentence and its surrounding context.

All participants were presented with the same pool of evaluation material and could evaluate up to all nine speaker test sets. Since participation was voluntary and participants could terminate the experiment at any time, not all participants completed the full questionnaire. We retained all completed judgments from incomplete questionnaires rather than discarding those participants. To reduce potential imbalance caused by incomplete responses, the order of the nine test sets was randomized independently for each participant. In addition, the order of the seven recordings within each test set was randomized. This randomization reduced systematic imbalance across speakers, synthesis tools, and spoofing conditions, but could not eliminate possible non-random dropout.

At the beginning of each speaker test set, participants were shown a reference video containing the bona fide voice of the speaker. This reference could be replayed at any point during the corresponding set. Participants then listened to the seven recordings in randomized order. After each recording, they evaluated each of the four sentence-level segments separately and indicated whether they perceived the segment as bona fide speech, synthetic speech, or whether they were uncertain. A synthetic segment was considered deceptive when a participant classified it as bona fide; the task did not test speaker identification. The position and time interval of each sentence-level segment were explicitly shown in the questionnaire to ensure that participants evaluated the intended part of the recording.

Before computing the evaluation metrics, uncertain answers were excluded from the analysis. The final dataset therefore consists only of binary authenticity judgments in which participants explicitly classified a segment as either bona fide or synthetic. Because incomplete questionnaires and excluded uncertain answers can lead to unequal numbers of valid judgments across speakers and conditions, the final number of valid responses was checked across synthesis tools and spoofing conditions before calculating aggregate performance metrics.

\subsection{Experiment setup}

Recruitment took place in person during the \event. To encourage engagement, the experiment included gamification elements, with participants awarded a material prize according to their overall accuracy. Participants received no feedback during the task and never learned whether an individual response was correct; their accuracy was revealed only after they completed the experiment.

During the listening sessions, participants were seated in a quiet room and listened to the recordings exclusively through headphones to minimize external distractions. Before the experiment began, participants were informed about the nature of the study and the data-handling procedures and subsequently provided informed consent. They then provided demographic information, including age group, gender, and occupation. The detailed assignment, randomization, and evaluation protocol is described in the previous \autoref{survey_structure}.

\subsection{Automated detector baselines}
\label{subsec:detectors}
To compare human judgments with automated countermeasures (RQ3), we evaluated six modern pretrained deepfake speech detectors on the same material presented to the participants. Recent comparative evaluation has shown that detector performance and robustness depend strongly on evaluation conditions and generalization to unseen samples~\cite{firc2025evaluation}. Each detector follows a common self-supervised architecture: a pretrained speech representation model as the frontend, followed by an anti-spoofing backend~\cite{stanek2026fusion}. We used two front-ends, XLS-R 300M~\cite{Babu2021} and WavLM Base+~\cite{wavlm}, and two back-ends, AASIST~\cite{jung2022aasist} and Multi-Head Factorized Attentive pooling (MHFA)~\cite{peng2022mhfa, rohdin24_asvspoof}. AASIST is a widely used graph-attention back-end~\cite{tak2022}, while MHFA is an attention-based pooling back-end adopted in recent anti-spoofing systems~\cite{rohdin24_asvspoof}. The systems differ in training data: two were trained on ASVspoof~2019~LA~\cite{WANG2020101114} and four on ASVspoof~5~\cite{wang2024asvspoof5}, spanning a range of architectures and data. Table~\ref{tab:detector_setup} summarizes the setup and each detector's Equal Error Rate (EER) on its source benchmark.

\begin{table}[t]
\centering
\caption{Automated detector baselines. EER is reported on the respective source benchmark used for training (eval subset of ASVspoof~2019~LA or ASVspoof~5), not on our study material.}
\label{tab:detector_setup}
\footnotesize
\begin{tabular}{lllc}
\toprule
Training data & Front-end & Back-end & EER \\
\midrule
\multirow{2}{*}{ASVspoof 2019 LA}
    & \multirow{2}{*}{XLS-R} & AASIST & 0.68\% \\
    &                        & MHFA   & 0.57\% \\
\midrule
\multirow{4}{*}{ASVspoof 5}
    & \multirow{2}{*}{XLS-R} & AASIST & 4.16\% \\
    &                        & MHFA   & 5.34\% \\
    \cmidrule(lr){2-4}
    & \multirow{2}{*}{WavLM} & AASIST & 4.06\% \\
    &                        & MHFA   & 5.26\% \\
\bottomrule
\end{tabular}
\end{table}

Each detector outputs a bona fide probability for a recording or segment. We binarized scores at a fixed threshold $t=0.5$ (scores $\geq 0.5$ interpreted as bona fide, lower as synthetic). We evaluated whole-recording predictions for full- and partial-spoof recordings, and, for partial spoofs, sentence-level predictions separately for manipulated and bona fide segments. Because this threshold is applied to out-of-distribution material and each per-condition estimate aggregates only nine speakers, we read the detector results as indicative rather than as a calibrated benchmark (see Section~\ref{sec:humans-vs-detectors}). These six systems are not intended to represent all contemporary detectors; rather, they provide controlled baselines spanning two front-ends, two back-ends, and two training datasets.

\subsection{Evaluation metrics}
\label{subsec:evaluation}

To assess the reliability of human judgments, we computed inter-rater agreement using Krippendorff's $\alpha$ on sentence-level authenticity labels ($\alpha = 1$ denotes perfect agreement, $\alpha = 0$ chance-level agreement, and values above ${\approx}\,0.8$ are conventionally treated as reliable). Krippendorff's $\alpha$ was selected because it accommodates nominal categories and incomplete response matrices, which is important in our setting, where participants could terminate the experiment at any time and uncertain responses were excluded from the analysis. Each item corresponded to one evaluated sentence-level segment, and responses were encoded as binary labels, bona fide or synthetic. Agreement was computed across all sentence-level judgments and separately for the main spoofing conditions and selected synthesis tools. Beyond these aggregate metrics, we report signal-detection measures with ``synthetic'' as the target and bona fide recordings as the noise distribution: sensitivity $d'$ (discrimination ability; $0$ = chance, higher = better) and criterion $c$ (response bias; $c<0$ indicates over-reporting ``synthetic'' and thus more false positives, $c>0$ the opposite). We compare full- versus partial-spoof detection per participant with the Wilcoxon signed-rank test, a paired non-parametric test that respects the clustering of judgments within listeners, and quantify uncertainty with cluster bootstrap $95\%$ confidence intervals resampled over participants.

To evaluate participants' responses, we use three metrics (\textit{F1 score, All OK, Over 50 percent}) with varying levels of rigor to provide a more detailed view of the results. Since it is not obvious which metric best captures the success of deepfake detection, we choose a combination of several. In this section, we present the metrics in order from the strictest to the loosest:

\textit{All OK} is the strictest metric, where a recording is classified as correct only if all four parts are identified without error. An error in a single segment means failure of the overall classification of the recording. This metric, calculated as the ratio of completely correctly evaluated recordings to their total number, penalizes even small errors, better reflecting the real challenges of detecting partial spoofs. In the context of disinformation, this approach is crucial because even overlooking a single manipulated sentence can fundamentally change the meaning of the entire message.

\textit{F1 score}~\cite{sasaki2007truth} is a metric used to evaluate the accuracy of models and decision processes. We use the F1 score to balance precision and recall, providing a robust measure for unbalanced datasets.

\textit{Over 50 Percent} is the most lenient metric, where classification is successful if at least three out of four parts of the recording are correctly identified. The resulting score represents the ratio of such successful evaluations to the total number of recordings. The metric recognizes that even partial recognition of synthetic elements, such as in vishing, serves as an important warning that increases the chance of detecting fraud.

However, this metric is unsuitable for partial spoofs, where only one part is changed. If a respondent marked the entire recording as bona fide, the metric would evaluate the answer as correct even though the synthetic segment was completely overlooked. This could skew the results and overestimate the participants' ability to detect manipulation.

In the first part of the experiment, we examine full spoof recordings where the entire audio is synthetic. We use the \textit{F1 score}, the \textit{Over 50 Percent} metric, and the \textit{All OK} metric. In the second part, we analyze partial spoof recordings with only one synthetic sentence. We use the \textit{F1 score} and the \textit{All OK} metric. 

Uncertain responses were treated as missing rather than as random guesses; the reported metrics therefore describe only definite binary judgments and may be biased if uncertainty differs across conditions.

For the automated baselines (Section~\ref{subsec:detectors}), we report the same \textit{F1 score} and \textit{All OK} metrics on the identical recordings, enabling a direct human-vs-detector comparison (RQ3).

\section{Results}

The experiment was conducted over three days at the \event event, with 82 volunteers participating. The full questionnaire comprised nine speaker test sets (63 recordings); on average participants completed close to all of them (mean 8.6 sets), spending about 35 minutes.

The experiment was conducted with a sample of IT professionals, selected based on the assumption that their familiarity with technology and higher level of critical thinking might enhance their ability to identify deepfake recordings or manipulated segments. However, performance in this technically literate sample cannot be assumed to represent performance in the general population.

Participants spanned a wide range of age groups: the largest were aged 35 to 44 (30) and 25 to 34 (27), followed by 18 to 24 (15) and 45 to 54 (10). Women were underrepresented (11 vs.\ 70 men, one ``other''), mirroring long-term gender trends in the IT sector~\cite{Du2019HourOC}. Although previous work suggests that gender plays only a minor role in deepfake detection ability~\cite{Malinka2024Comprehensive}, this imbalance limits demographic generalizability.

\begin{table}[t]
\centering
\caption{Inter-rater agreement for sentence-level authenticity judgments. Uncertain and missing responses were excluded before computing Krippendorff's $\alpha$. Items denote unique sentence-level segments.}
\label{tab:inter_rater_agreement}
\footnotesize
\begin{tabular}{lrrr}
\toprule
Condition & Items & Judgments & $\alpha$ \\
\midrule
All sentence-level judgments & 252 & 17,233 & 0.631 \\
\midrule
Full spoof & 108 & 7,392 & 0.502 \\
Partial spoof & 108 & 7,429 & 0.528 \\
\midrule
Full spoof, YourTTS & 36 & 2,569 & 0.035 \\
Full spoof, RTVC & 36 & 2,514 & 0.011 \\
Full spoof, ElevenLabs & 36 & 2,309 & 0.155 \\
\midrule
Partial spoof, mixYourTTS & 36 & 2,519 & 0.639 \\
Partial spoof, mixRTVC & 36 & 2,491 & 0.593 \\
Partial spoof, mixElevenLabs & 36 & 2,419 & 0.120 \\

\bottomrule
\end{tabular}
\end{table}

After questions about demographic data, respondents moved on to the main part of the experiment. We now focus on the analysis of the results, in which the ability of respondents to recognize synthetic voice recordings is assessed.

Before addressing the individual research questions, we first evaluate the reliability of the collected human judgments. Table~\ref{tab:inter_rater_agreement} reports the inter-rater agreement results. Across all sentence-level judgments, agreement was moderate, with Krippendorff's $\alpha = 0.631$. This indicates that participant responses were not random or purely idiosyncratic. Agreement was also moderate when aggregating full-spoof and partial-spoof conditions separately, with $\alpha = 0.502$ and $\alpha = 0.528$, respectively.

Two patterns require careful interpretation. First, the full-spoof old-tool conditions show near-zero agreement (RTVC $\alpha = 0.011$, YourTTS $\alpha = 0.035$) despite $\approx$90\% accuracy. This is the well-known \emph{prevalence (base-rate) paradox} of chance-corrected agreement, not genuine disagreement: when the marginals are highly skewed (here almost every listener correctly and near-unanimously labels these obvious fakes as synthetic), the expected (chance) agreement approaches the observed agreement, so $\alpha$ collapses even though raw percent agreement exceeds 90\%. Krippendorff's $\alpha$ therefore measures consistency, not correctness, and must be read jointly with the accuracy metrics below. Second, and more meaningfully, agreement is low for ElevenLabs precisely where accuracy is \emph{also} near chance and the marginals are close to balanced (full-spoof $\alpha = 0.155$, partial-spoof $\alpha = 0.120$): here low agreement does reflect genuine perceptual ambiguity, as listeners disagree because no reliable cue exists. The contrast within partial spoofs is telling: agreement is high for the audibly flawed older tools (mixRTVC $\alpha = 0.593$, mixYourTTS $\alpha = 0.639$) but collapses for mixElevenLabs.

Based on the captured responses and their evaluation, we are able to answer the research questions:

%SEM POJDU ODPOVEDE NA JEDNOTLIVE RESEARCH QUESTION
\noindent \textbf{RQ1:} \textit{How does human voice-deepfake detection performance differ among the selected speech synthesizers?}

\begin{figure}[t]
    \centering
    \includegraphics[width=0.84\linewidth]{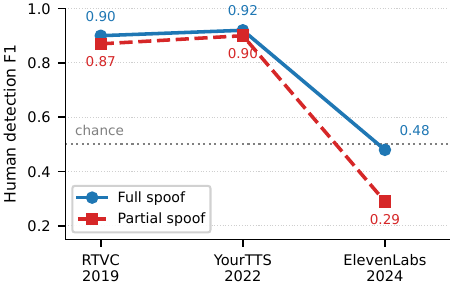}
    \caption{Human detection performance (mean F1) for full and partial spoofs from the three selected synthesizers. Detection remains high for RTVC and YourTTS and is substantially lower for ElevenLabs. The dotted line marks an F1 reference value of 0.5.}
    \label{fig:trend}
\end{figure}

\begin{figure*}[t]
    \centering
    \includegraphics[width=0.95\linewidth]{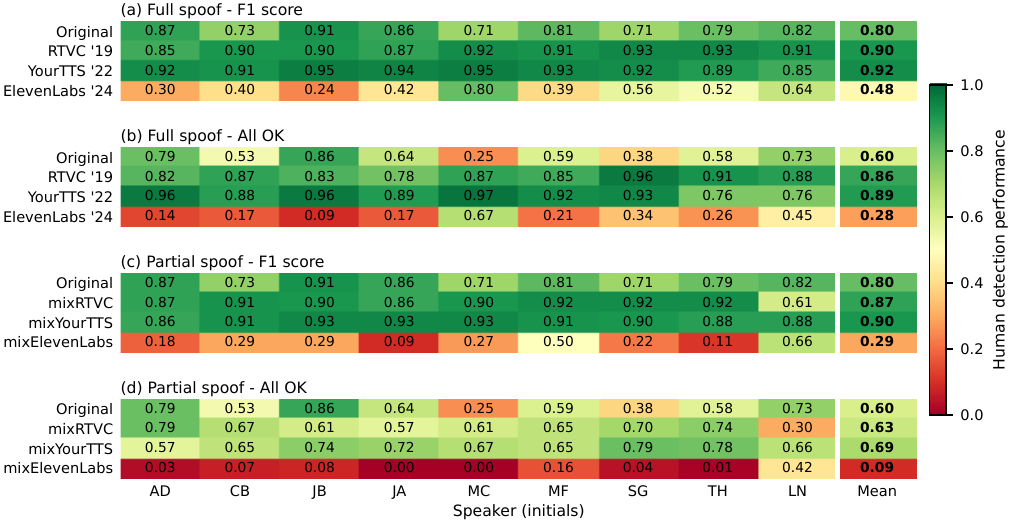}
    \caption{Human detection per speaker for (a,b) \emph{full-spoof} and (c,d) \emph{partial-spoof} recordings (the prefix ``mix'' marks one synthetically replaced sentence). Rows: recording source (bona fide Original, RTVC~2019, YourTTS~2022, ElevenLabs~2024); columns: speaker initials (AD~=~Adele, CB~=~Cate Blanchett, JB~=~Jeff Bridges, JA~=~Jennifer Aniston, MC~=~Miley Cyrus, MF~=~Morgan Freeman, SG~=~Selena Gomez, TH~=~Tom Holland, LN~=~Liam Neeson), with the final \emph{Mean} column. Green~=~reliably detected, red~=~deceptive. Lower scores on the \emph{Original} row reflect false positives (bona fide speech judged synthetic); the \emph{mixElevenLabs} rows (c,d) show the partial-spoof collapse.}
    \label{fig:heatmaps}
\end{figure*}

Figure~\ref{fig:trend} summarizes the central finding: human recognition of fully synthetic speech remains high for RTVC and YourTTS but is substantially lower for the selected ElevenLabs recordings. Figure~\ref{fig:heatmaps}(a,b) breaks the result down per speaker as heat maps, where rows denote the recording source in chronological order (bona fide Original, RTVC, YourTTS, ElevenLabs), columns denote speaker initials, and the final \emph{Mean} column aggregates across the nine speakers.

The results, presented in Fig.~\ref{fig:heatmaps}(a,b), reveal a wide range of observed scores. Notably, detection of recordings generated using ElevenLabs, the most recent tool, drops sharply relative to the older tools. When requiring respondents to correctly classify all four parts of a recording (\textit{All OK}), the mean success rate was only 28\%; the more lenient \textit{Over 50 Percent} metric reached 33\%, and the \textit{F1 score} averaged 48\%, down from $\approx$0.90 for the older tools. The chance levels make the collapse concrete. Each sentence is a binary bona fide/synthetic decision, so a guesser scores 50\% per sentence; the \textit{Over 50 Percent} rate (correct on at least three of four sentences) then has a $\approx$31\% chance level and the \textit{All OK} rate (all four correct) a $6.25\%$ chance level. The observed \textit{Over 50 Percent} of 33\% therefore sits essentially at its chance level, and although 28\% \textit{All OK} exceeds its much stricter 6.25\% baseline, it still corresponds to near-random per-sentence behavior. Indeed, mean per-participant per-sentence accuracy on full-spoof ElevenLabs is only 0.43 (95\% CI [0.37, 0.50], cluster bootstrap over participants), at or just below the 0.5 chance level, versus 0.96 to 0.97 for the older tools. (We treat F1 as a descriptive balance of precision and recall rather than as a chance-referenced score, since its baseline depends on prevalence.) ElevenLabs thus generates voices that are far harder to detect than those of older models.

With older tools (RTVC, YourTTS), respondents achieve an average F1 score and a strict All OK metric of approximately 90\%. These results, consistent with the existing literature, confirm that synthetic speech from older technologies is still reliably recognizable to humans. Notably, recognition does not degrade monotonically with release year: YourTTS (2022) is detected as well as, or slightly better than, RTVC (2019). With only three selected systems, our data therefore characterizes system-specific differences rather than a general temporal trajectory. In contrast, the selected ElevenLabs recordings pose a serious security risk because their quality makes detection based on voice features substantially more difficult. Listeners may therefore accept synthetic speech as bona fide under these conditions, which undermines the credibility of voice communication and highlights the need for continued monitoring of deepfake technologies.

This collapse has a clear implication: as synthetic speech becomes more natural, human perception alone is no longer a reliable defense, and safeguards beyond perceptual judgment are urgently needed. Automated detection is not yet a dependable fallback either, however, as we show in Section~\ref{sec:humans-vs-detectors}.

The data also reveal a counterintuitive pattern. For the older tools, bona fide recordings are judged \emph{less} reliably than clearly artificial synthetic ones: the \emph{Original} row in Fig.~\ref{fig:heatmaps} scores only $0.60$ (All~OK), compared with $0.86$--$0.89$ for the synthetic samples. A signal-detection analysis clarifies why. Treating ``synthetic'' as the target and the bona fide \emph{Original} segments as the noise distribution, listeners are markedly \emph{liberal}: quick to label audio as ``fake''. This is reflected by a $17.5\%$ false-alarm rate on bona fide clips. At the same time, sensitivity $d'$ differs sharply across the selected systems, from $2.75$ (RTVC) and $2.88$ (YourTTS) to just $0.73$ for ElevenLabs; discrimination is not eliminated, but it approaches the $d'=0$ floor. The two effects combine in a striking way: despite the liberal tendency, most ElevenLabs fakes are convincing enough to fall below the listener's threshold for ``fake'' and are accepted as bona fide (hit rate $0.42$), pushing full-spoof ElevenLabs accuracy to $0.43$, at or below the $0.5$ chance level. Because aggregate accuracy thus conflates sensitivity with response bias, we treat the full-versus-partial contrast and the ElevenLabs $d'$ collapse as our central findings, and we read the $17.5\%$ false-alarm rate on bona fide audio as a direct measure of the over-rejection of real recordings---an erosion of trust we revisit in the discussion.

\noindent \textbf{RQ2:} \textit{How does the human ability to recognize voice deepfake change when deepfakes are embedded between bona fide recordings? }

The results of the second part of the experiment show a strong correlation to the first part. As seen in Fig.~\ref{fig:heatmaps}(c,d), most respondents fairly easily recognized the recordings synthesized using older tools, with a mean F1 score of almost 90\%. In contrast, recordings produced with ElevenLabs again yield markedly lower F1 and All~OK scores: the mean F1 for mixElevenLabs is only 29\%, and under the strictest \textit{All OK} criterion success falls below 10\%.

When comparing the two parts of the experiment, partial-spoof recordings deceive respondents even more than full spoofs, with the most striking difference observed for ElevenLabs (compare Fig.~\ref{fig:heatmaps}(a,b) with (c,d)). For ElevenLabs, embedding the synthetic sentence in bona fide speech lowers the mean F1 by roughly $1.7\times$ (from 0.48 to 0.29) and the strict All~OK rate by roughly $3\times$ (from 0.28 to 0.09); a smaller decrease is visible for the other tools.

We test this rigorously while respecting that each listener contributes many correlated judgments: for each participant we compute the detection rate of the synthetic segment(s) in full- versus partial-spoof recordings of the same generator, and compare the paired per-participant rates with a Wilcoxon signed-rank test. Detection was significantly lower in the partial-spoof condition for every generator. The difference was largest for ElevenLabs, where detection of the fake fell from 0.43 (full) to 0.23 (partial) ($n=73$ paired participants, $p\approx6\times10^{-8}$); it was smaller but still significant for RTVC ($0.96\rightarrow0.90$, $p\approx2\times10^{-6}$) and YourTTS ($0.97\rightarrow0.94$, $p=0.02$). Because the full- and partial-spoof conditions also differed in linguistic content and surrounding context, this comparison does not isolate the causal contribution of bona fide context alone.

\subsection{Error analysis.}
To better understand why partial-spoof recordings were more difficult to detect than full-spoof recordings, we analyzed the types of classification errors made by participants at the sentence level. In particular, we distinguished between errors on bona fide segments and errors on manipulated segments. This allows us to determine whether the decrease in performance was caused mainly by participants falsely marking bona fide speech as synthetic, or by accepting synthetic speech as bona fide.

The results show that the dominant error pattern differs across synthesis tools. For older tools, RTVC and YourTTS, participants were still able to identify both the manipulated and bona fide segments with relatively high accuracy. As shown in Table~\ref{tab:paritialdeepfake}, the accuracy for detecting the deepfake sentence reached 90.95\% for RTVC and 94.26\% for YourTTS, while the accuracy for bona fide sentence segments remained close to 90\%. This indicates that, for older synthesis systems, participants could still perceive artifacts in the manipulated sentence and separate it from the surrounding bona fide speech.

The pattern changes substantially for ElevenLabs. In partial-spoof recordings generated with ElevenLabs, participants correctly identified the manipulated sentence in only 22.65\% of cases, while bona fide sentence segments were still classified correctly in 88.59\% of cases. This shows that the decline in performance was not caused by a general inability to follow the task or by random responding. Instead, participants were specifically unable to detect the synthetic sentence when it was embedded in an otherwise bona fide recording. The bona fide context therefore appears to increase the perceived credibility of the manipulated segment.

Partial spoofing thus poses a qualitatively different challenge than full spoofing. In full spoofs, listeners judge the whole recording on global voice quality, prosody, and artifacts; in partial spoofs, the manipulated sentence is surrounded by bona fide speech from the target speaker, which may reduce suspicion and make a short high-quality insert appear consistent with the rest of the utterance. The main risk, then, is not uniform confusion but the targeted acceptance of a single synthetic sentence in a credible context. This is precisely the regime of political statements, legal evidence, media interviews, and voice phishing, where one altered sentence can change a message's meaning. This effect appears for RTVC and YourTTS too, but far more weakly, indicating that the selected ElevenLabs samples both mimic the voice more realistically and mask the synthetic segment more effectively within bona fide speech.

\subsection{Humans versus automated detectors}
\label{sec:humans-vs-detectors}
\begin{figure*}[t]
    \centering
    \includegraphics[width=0.92\linewidth]{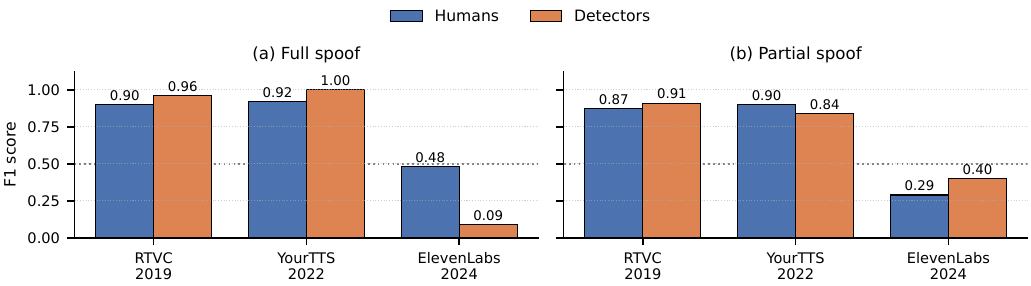}
    \caption{Humans versus six pretrained automated detectors (mean F1) on the same recordings, grouped by synthesizer. On ElevenLabs, detectors (0.09) fall below humans (0.48) on full spoofs (panel a), whereas on partial spoofs (panel b) detectors (0.40) exceed humans (0.29), though their strict All~OK localization rate is 0.00. Detector scores are averaged across all six detectors.}
    \label{fig:human_vs_detector}
\end{figure*}

To position these human results within the broader anti-spoofing landscape, we evaluated the six detectors of Section~\ref{subsec:detectors} on the same material; Fig.~\ref{fig:human_vs_detector} compares the two. The two defenses fail in \emph{complementary} ways. For RTVC and YourTTS, the selected detectors are highly reliable (mean F1 $0.96$ for RTVC and $1.00$ for YourTTS on full spoofs), comfortably exceeding human performance. On the modern commercial tool, however, the full-spoof picture inverts: detectors collapse to a mean F1 of only $0.09$, substantially \emph{worse} than humans ($0.48$), a result consistent with domain mismatch between the detector training data and the selected ElevenLabs samples. Partial spoofing reverses the ranking again: on mixElevenLabs the detectors reach mean F1 $0.40$, above the humans' $0.29$. Yet this apparent advantage is illusory under strict scoring, because the detectors' \textit{All OK} rate on mixElevenLabs is $0.00$ (versus $0.09$ for humans), so \emph{neither} party reliably localizes the manipulated sentence. In short, humans collapse on partial spoofs where detectors hold up on whole-recording F1, detectors collapse on modern full spoofs where humans retain residual ability, and both fail the sentence-level localization that partial spoofing demands.

Applied to individual sentence segments in partial-spoof recordings, the detectors localize the manipulated sentence with detection rates of 0.519 to 0.778 and accept bona fide segments at rates of 0.630 to 1.000. Fine temporal resolution thus helps, but performance stays inconsistent across architectures and training conditions, reinforcing the need for methods designed to localize short manipulated segments rather than classify whole recordings.

\begin{table}[t]
    \centering
    \caption{Human accuracy of identifying individual sentences based on their type for \textit{partial spoof} recordings.}
    \label{tab:paritialdeepfake}
    \begin{tabular}{lcc}
        \toprule
        \textbf{Model} & \textbf{Deepfake part (\%)} & \textbf{bona fide part (\%)} \\
        \midrule
        mixRTVC        & 90.95 & 88.07 \\
        mixYourTTS     & 94.26 & 90.13 \\
        mixElevenLabs  & 22.65 & 88.59 \\
        \bottomrule
    \end{tabular}
\end{table}

\section{Discussion}

\subsection{Why human detection collapses}
Detection tracks synthesis quality: older tools (RTVC, YourTTS) leave audible artifacts such as unnatural intonation, robotic timbre, and dropouts, which listeners exploit (and the $d'$ analysis above confirms high sensitivity), whereas ElevenLabs removes these cues to near indistinguishability ($d'=0.73$). The decisive point is that this collapse occurs \emph{even though participants were explicitly told deepfakes were present}: where Malinka et al.~\cite{Malinka2024Comprehensive} found particularly low performance among unaware listeners, we find low detection for the selected ElevenLabs samples even under an explicit-warning condition. Without an unwarned control group, however, we cannot determine how much the warning helped.

The qualitative responses explain \emph{why}. We coded the free-text comments of the 72 of 82 participants who left such comments, and grouped participants into quartiles by their All~OK score. Lower-performing listeners (Q1, Q2) relied on coarse impressions (``robotic voice'', ``something is off''), while top performers (Q3, Q4) cited specific irregularities in pitch, cadence, emotional flatness, and audio artifacts. For older tools this expertise paid off (mean All~OK reaches 0.89 to 0.94 for Q4 versus 0.23 to 0.24 for Q1). For the selected ElevenLabs samples, the advantage narrows: even the best listeners score 0.08 on All OK (Q4) versus 0.01 (Q1), suggesting that the cues used for the older tools were less useful in this condition.

\subsection{Partial spoofing: the most deceptive threat}
Replacing a single sentence in an otherwise bona fide utterance is markedly more deceptive than full synthesis. The error analysis (Table~\ref{tab:paritialdeepfake}) localizes the effect: for ElevenLabs, listeners do not fail globally but \emph{specifically} on the synthetic insert (detected only 22.65\% of the time), while still judging bona fide segments correctly (88.59\%). This pattern is consistent with the surrounding bona fide speech increasing the perceived credibility of the manipulated segment, although the design does not isolate context from linguistic differences. This is precisely the regime exploited in disinformation, fabricated evidence, and voice phishing, where altering one sentence changes the meaning of an entire message. Our participants are verified IT professionals with above-average technological literacy and risk awareness~\cite{Torten2018, Mersinas2015}; nevertheless, the findings characterize this population only and cannot be treated as an upper bound for the general public.

\subsection{Humans, detectors, and the limits of each}
The two defenses fail in complementary ways: detectors are strong on the synthesis they were trained on but fall below humans on out-of-distribution full-spoof ElevenLabs (0.09 vs.\ 0.48), whereas humans keep some residual ability there yet perform poorly on partial spoofing, and neither localizes the manipulated sentence (strict All~OK $\leq$0.09). No single line of defense suffices against modern, partial, or commercial synthesis, which argues for layered defenses rather than reliance on humans or any one detector.

\subsection{Security implications}
Declining human detection amplifies the risk of voice phishing and disinformation; combined with large language models, deepfake voice enables scalable, hard-to-defend social engineering~\cite{vishing,wenger2021hello}. Synthetic speech also poses a demonstrated threat to speaker-recognition mechanisms in consumer voice assistants~\cite{malinka2024resilience}, while partial spoofing magnifies the broader threat by lending an attack the credibility of bona fide audio. A subtler harm is the rise in false positives: as listeners grow wary, they increasingly reject \emph{bona fide} recordings, the so-called liar's dividend~\cite{chesney2019deepfakes}, eroding trust in telephone communication, voice authentication, journalism, and audio as legal evidence.

\subsection{Mitigation}
No single measure suffices; defenses must span technical, educational, and regulatory domains. \emph{Technically}, recording-level detectors are inadequate: detection must localize short manipulated segments~\cite{zhang2023partialspoof, zhang2025partialedit, zhang2024spoofdiar, wu2024coarse,liu2024harmonet} and generalize to out-of-distribution generators, since ours failed on unseen commercial synthesis. Because detection alone is brittle, proactive \emph{provenance} is essential: content-authenticity standards such as C2PA~\cite{c2pa2025spec} and robust watermarking embedded during generation~\cite{sanroman2024audioseal} enable a verifier to confirm authenticity without relying on perceptible artifacts. \emph{Educationally}, training must move beyond ``listen for artifacts.'' In our experimental condition, the selected high-quality synthetic samples provided few reliable cues, and warning alone did not yield high detection performance. \emph{Regulatory} approaches may also consider disclosure and watermarking requirements for synthetic media.

\subsection{Limitations}
Our sample is restricted to IT professionals, so the findings should be validated on broader, non-expert populations. Participants knew deepfakes might be present, which likely increased both suspicion and false positives compared with an unprimed setting. We excluded ``uncertain'' responses before scoring, so the reported metrics are conditional on definite judgments and may be biased if uncertainty differs across conditions. Voluntary early termination may make dropout non-random, though order randomization and the balanced per-condition judgment counts (Table~\ref{tab:inter_rater_agreement}) mitigate this. The fixed high prevalence of synthetic recordings may also have increased participants' tendency to answer ``synthetic.'' We did not measure participants' prior familiarity with the selected celebrities, although a reference video was provided at the beginning of each speaker set. We also did not stratify detector performance by speaker characteristics, although such characteristics can significantly affect detector performance across sex, language, age, and synthesizer type~\cite{stanek2025scdf}. Because generated replacement text was used for partial spoofs, acoustic effects cannot be completely separated from linguistic and contextual effects. Only one synthesizer represents each release period, and ElevenLabs is also the only commercial system; the observed differences therefore do not establish a general temporal trend. The survey did not collect per-question timestamps or response-time intervals, which limits our ability to analyze decision latency during the task. Reaction times, learning and fatigue effects, and demographic moderators are left to future work, as is a fuller detector benchmark beyond the selected fixed-threshold pool used here.

\subsection{Ethical considerations}
The study was approved by the ethics committee at Brno University of Technology with a decision number EK:2024-002. Participants gave informed consent, were fully anonymized, and could withdraw at any time. To minimize psychological risk and reputational damage to selected public figures, the generated statements were designed to avoid sensitive, political, personal, or offensive language.

\subsection{Data availability}

The survey audio dataset is not publicly available because it contains deepfake speech from identifiable public figures, but it may be provided upon reasonable request.

\section{Conclusion}
Across the three selected synthesizers, human voice-deepfake detection was substantially lower for the ElevenLabs samples than for RTVC and YourTTS, reaching near-chance per-sentence accuracy for full spoofs and below 10\% under the strict All~OK metric for partial spoofs, \emph{even when listeners were warned and technically skilled}. Partial spoofing was particularly difficult: a single synthetic sentence in bona fide speech was classified as bona fide 77\% of the time, and the selected detectors did not consistently close the gap. Because only one system represents each release period, these results demonstrate system-specific differences rather than a general generational trend. The key takeaways are: (i) warning alone did not yield high detection performance for the selected ElevenLabs samples, although the effect of warning was not independently tested; (ii) partial spoofing challenges both humans and recording-level detectors, motivating segment-localized methods; and (iii) the findings support considering layered defenses, including procedural verification, provenance, watermarking, and continuous detector retesting.

% % --------------------------------------------------------
\section{Acknowledgements}
This work was supported by the Brno University of Technology internal project FIT-S-26-9011.

% % --------------------------------------------------------
\section{Generative AI Use Disclosure}

During the preparation of this work, the authors used Generative AI Models (specifically Google Gemini, ChatGPT, and Grammarly) for language editing and text refinement. The authors reviewed and edited the output as needed and take full responsibility for the publication's content.

\bibliographystyle{IEEEtran}
\bibliography{mybib}

\end{document}